\documentclass[prb,twocolumn,superscriptaddress,showpacs,amsmath]{revtex4}
\usepackage{graphicx,color}
\usepackage{CJK}
\usepackage{bm}
\usepackage[hypertex]{hyperref}
\usepackage{float}

\begin{document}

\title{Majorana dc Josephson current mediated by a quantum dot}

\author{Luting Xu}
\affiliation{Center for Advanced Quantum Studies and
Department of Physics, Beijing Normal University,
Beijing 100875, China}

\author{Xin-Qi Li}
\email[]{lixinqi@bnu.edu.cn}
\affiliation{Center for Advanced Quantum Studies and
Department of Physics, Beijing Normal University,
Beijing 100875, China}

\author{Qing-Feng Sun}
\email[]{sunqf@pku.edu.cn}
\affiliation{International Center for Quantum Materials, School of Physics, Peking University, Beijing 100871, China}
\affiliation{Collaborative Innovation Center of Quantum Matter, Beijing 100871, China}

\begin{abstract}
The Josephson supercurrent through the hybrid
Majorana--quantum dot--Majorana junction is investigated. We particularly analyze the effect of spin-selective coupling
between the Majorana and quantum dot states, which emerges only in the topological phase and
will influence the current through bent junctions and/or
in the presence of magnetic fields in the quantum dot.
We find that the characteristic behaviors
of the supercurrent through this system
are quite counterintuitive, remarkably
differing from the resonant tunneling,
e.g., through the similar (normal phase)
superconductor--quantum dot--superconductor junction.
Our analysis is carried out under the influence of full set-up
parameters and for both the $2\pi$ and $4\pi$ periodic currents.
The present study is expected to be relevant to
future exploration of applications of the
Majorana-nanowire circuits.
\end{abstract}
\maketitle

\section{Introduction}

Majorana fermions (MFs) are exotic self-Hermitian particles
with non-Abelian statistics, hold a property with
themselves as their own antiparticles \cite{Mj37,Wil09,Franz10}
and promise robust building blocks for
topological quantum computation \cite{DS08,Fra15}.
Remarkable insight predicts that MFs can emerge as
novel excitations of Majorana zero modes
or Majornana bound states
in condensed matter systems,
e.g., from the non-Abelian excitations
in a 5/2 fractional quantum Hall effect
in semiconductor heterostructures \cite{Re91},
and based on exotic superconductors where MFs
correspond to zero-energy states of
an effective Bogoliubov-de Gennes Hamiltonian \cite{Kit01,Iva01}.

More recent proposals employ the proximity effect
from a conventional superconductor,
either in nanowires in the presence of
strong spin-orbit interaction and Zeeman spliting
\cite{Deng12,Fink13,Lee14,Mourik12,Lut10,Oppen10,Xu14},
or in topological insulators
\cite{Fu08,Bee09,Chiu11,Xu14b,Xu15}.
These efforts bring the MFs closer to experimental realization
and predict more reliable experimental signatures of their presence.
Among the signatures include such as the half-integer
conductance quantization \cite{Wim11},
the zero-bias peak in the tunneling conductance
\cite{add1,add2,add3,add4}, and the $4\pi$ Josephson
effect in superconductor-superconductor junctions
\cite{Lut10,Fu09,Ali11,ZSF1}.
In particular, in order to distinguish MFs from
other quasi-particle states,
some interest in recent years turns to the
{\it spin selective} Andreev reflection
\cite{Ng14,Oreg15,Hu15,Jia16}.

\begin{figure}[H]
  \centering
  \includegraphics[scale=0.6]{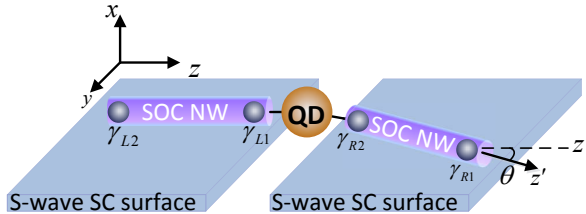}\\
  \caption{(color online)
Set-up sketch of the Majorana--QD--Majorana
junction, realized by semiconductor nanowires
in contact with the $s$-wave superconductors.
The two nanowires may have a mutual orientation angle
which is expected to affect the supercurrent
via the unique spin-selective coupling
between the Majorana and QD states.  }\label{system}
\end{figure}

In this work we consider the hybrid system
of Majorana--quantum dot--Majorana junction
which can be realized from semiconductor nanowires
in proximity-contact with $s$-wave superconductors,
as schematically shown in Fig.~\ref{system}.
Similar systems of Majorana nanowire coupled to
quantum dots (QDs) have been investigated
for phenomena such as teleportation \cite{Lee08},
anomaly of conductance peak \cite{Baranger11},
characteristic signatures in current noise spectrum \cite{li12},
and featured Josephson current \cite{Xu14}.
Our present interest is the dc Josephson current
under the influence of spin-selective
coupling between the Majorana and QD states,
which is most relevant to {\it bent junctions}
and the presence of magnetic field in the QD area.

Due to the helical property of MF,
the MF at the end of the nanowire only
couples to a unique spin state in the normal region,
e.g., the spin-up QD state as shown in Fig.~\ref{system}.
Actually, this spin-selective coupling is
the origin of the spin-dependent Andreev reflection \cite{Lopez13,Ng14}.
In previous studies,
the set-up configuration is usually assumed to be either
a single Majorana nanowire coupled to normal leads or QDs,
or a straight Majorana--normal region--Majorana
junction with bent angle $\theta=0$.
In both cases, only the spin-up states (in the normal parts)
couple to the MF and contribute to the current.
However,
if the Majorana--QD--Majorana junction is not straight
(with a bent angle $\theta\not=0$ between the nanowires),
the two MFs at the ends of the nanowires (see Fig.~\ref{system})
will couple to spin states in the QD with different orientations,
leading thus to both the spin-up and spin-down states
participating in the transport.
This is anticipated to result in different
current behaviors of the straight and bent junctions.
On the application aspect
(e.g. in the topological quantum computations),
the Majorana nanowires will possibly
have an orientational angle,
or one would like to employ this orientational angle
to modulate the charge transfer properties.
This makes thus the bending structure
studied in this work relevant to possible real circuits.  

We will also analyze the effect
of magnetic fields in the QD area.
This is motivated by viewing that,
in order to induce the emergence of MF,
magnetic fields are needed to apply in the nanowires,
which must spill over in the QD area
owing to its close separations from the nanowires.
Similar to the consequence of junction bending, we expect that
the non-$z$-axial direction magnetic field
will make the spin-down state involved in the transport as well.
We will show that, remarkably,
owing to the spin-selective coupling,
both the magnetic field in the QD area
and the junction bending will result in
some counterintuitive behaviors of the Josephson current.
For instance, in the case of QD level aligned with
the Fermi energy (under resonant tunneling),
the Josephson current is to be strongly suppressed
by the magnetic field and junction bending.
However, as the QD level deviates from the Fermi level
(violates the resonant-tunneling condition),
the oscillation amplitude of the Josephson current
always shows an enhanced value,
together with robust jumps in the current-phase curves.

\section{Model and Methods}

\subsection{Set-up Model}

In this work we consider the simple setup as shown in Fig.~\ref{system},
where two semiconductor nanowires are connected through a QD.
The nanowires are in proximity contacted with an $s$-wave superconductor.
Then, the proximity-effect-caused superconductivity,
together with the strong Rashba spin-orbit coupling (SOC)
and Zeeman splitting inside the nanowires,
can possibly induce emergence of a pair of MFs at the ends of each nanowire \cite{Lut10,Oppen10},
as denoted in Fig.~\ref{system} by $\gamma_{L\, 1,2}$
for the left wire and $\gamma_{R\, 1,2}$ for the right one.

The tunnel-coupled system can be modeled by an effective
low-energy Hamiltonian $H=H_M+H_{dot}+H_{dL}+H_{dR}$,
more explicitly with
\begin{subequations}
\begin{align}
H_M&=i\varepsilon_L\gamma_{L1}\gamma_{L2}
+i\varepsilon_R\gamma_{R1}\gamma_{R2}\;,\\
H_{dot}&=\sum_{\sigma}\varepsilon_d d^\dag_{\sigma}d_{\sigma}+(d^\dag_\uparrow,d_\downarrow^\dag)
\vec{\sigma}\cdot\vec{B}
\begin{pmatrix}
d_\uparrow\\
d_\downarrow
\end{pmatrix}\;,\\
H_{dL}&=(\lambda_L d_\uparrow
-\lambda_L^*d^\dag_\uparrow)\gamma_{L1}\;,\\
H_{dR}&=i(\lambda_R\tilde{d}_\uparrow
+\lambda_R^*\tilde{d}^\dag_\uparrow)\gamma_{R2} \;.
\end{align}\label{Ham}
\end{subequations}
Here $H_M$ is the effective low energy Hamiltonian
for the two pairs of Majorana states
emerged at the ends of the two nanowires.
Each Majorana pair may have nonzero coupling energy,
i.e., $\varepsilon_{L}\sim e^{-l_{L}/\xi_{L}}$
and $\varepsilon_{R}\sim e^{-l_{R}/\xi_{R}}$,
where $l_{L(R)}$ and $\xi_{L(R)}$ are, respectively,
the length of the nanowire and
the superconductor coherence length.
$H_{dot}$ denotes the QD Hamiltonian,
with a single spatially quantized level $\varepsilon_d$
(tunable by gate voltage),
and possibly affected via the Zeeman effect
by magnetic field $\textbf{B}=(B_x,B_y,B_z)$
 (Here we assume an arbitrary direction of magnetic field).
$d^\dag_\sigma$ and $d_\sigma$ are the creation
and annihilation operators of the QD electron with spin $\sigma$.

$H_{dL}$ and $H_{dR}$ describe the tunnel coupling
between the dot and the nearby Majorana states.
In general, the coupling amplitudes can be expressed as
$\lambda_{L(R)}=|\lambda_{L(R)}|e^{i\phi_{L(R)}/2}$,
where the phase factors are determined by the phase of the substrate
superconductors and their difference will result in
the famous Josephson current.
Another important issue to be noted is that, due to the helical property of MF,
the MF only couples to the spin-up state in the QD
(defined in the same $z$-representation of the associated nanowire)
\cite{Lopez13,Ng14}.
As mentioned in the introduction, our special interest in this work
is to consider the two nanowires not aligned in the same
orientation, but with an angle $\theta$, as shown in Fig.\ref{system}.
Thus, the left and right MFs would couple only to the QD state spin-polarized,
respectively, along the $z$ and $z^\prime$ axes,
with the associated electron operators
connected by the following unitary transformation:
\begin{eqnarray}
\begin{pmatrix}
\tilde{d}_\uparrow\\
\tilde{d}_\downarrow
\end{pmatrix}
=\begin{pmatrix}
\cos{\frac{\theta}{2}}&\sin{\frac{\theta}{2}}\\
-\sin{\frac{\theta}{2}}&\cos{\frac{\theta}{2}}
\end{pmatrix}
\begin{pmatrix}
d_\uparrow\\
d_\downarrow
\end{pmatrix}  \,.
\end{eqnarray}
This implies that, if the bent angle $\theta\not=0$,
the spin-down state also couples to the MFs.

In practice, it would be convenient to convert the MFs
to regular fermion representation, via the simple transformation:
$c_L=(\gamma_{L1}+i\gamma_{L2})/\sqrt{2}$,
and $c_R=(\gamma_{R1}+i\gamma_{R2})/\sqrt{2}$.
The creation operators are their Hermitian conjugate
and satisfy $\{c_{L(R)},c_{L(R)}^\dag\}=1$.
Now let us apply the generalized Nambu representation,
by introducing the field creation and annihilation
operators as
$\psi^\dag=(d^\dag_\uparrow,d^\dag_\downarrow,
d_\downarrow,d_\uparrow,c_L^\dag,c_L,c_R^\dag,c_R)$,
and $\psi=(d_\uparrow,d_\downarrow,d^\dag_\downarrow,
d^\dag_\uparrow,c_L,c_L^\dag,c_R,c_R^\dag)^T$.
Then, the Hamiltonian of the whole system can be rewritten as
\begin{equation}
H=\frac{1}{2}\psi^\dag \mathcal{H}\psi  \,,
\end{equation}
where the Hamiltonian matrix has a block form given by
\begin{eqnarray}
\mathcal{H}=\begin{pmatrix}
\mathcal{H}_{dd}&\mathcal{H}_{dL}&\mathcal{H}_{dR}\\
\mathcal{H}_{Ld}&\mathcal{H}_{LL}&0\\
\mathcal{H}_{Rd}&0&\mathcal{H}_{RR}
\end{pmatrix}   \,,
\end{eqnarray}
and each sub-matrix reads, respectively,
\begin{subequations}
\begin{gather}
\mathcal{H}_{dd}=\begin{pmatrix}
\varepsilon_d+B_z&B_x-iB_y&0&0\\
B_x+iB_y&\varepsilon_d-B_z&0&0\\
0&0&-\varepsilon_d+B_z&-B_x+iB_y\\
0&0&-B_x-iB_y&-\varepsilon_d-B_z
\end{pmatrix}  \,, \\
\mathcal{H}_{LL}=\begin{pmatrix}
\varepsilon_L&0\\
0&-\varepsilon_L
\end{pmatrix}   \,,  \\
\mathcal{H}_{RR}=\begin{pmatrix}
\varepsilon_R&0\\
0&-\varepsilon_R
\end{pmatrix}   \,, \\
\mathcal{H}_{dL}=\frac{1}{\sqrt{2}}\begin{pmatrix}
-\lambda_L^*&-\lambda_L^*\\
0&0\\
0&0\\
\lambda_L&\lambda_L
\end{pmatrix}   \,, \\
\mathcal{H}_{dR}=\frac{1}{\sqrt{2}}\begin{pmatrix}
\lambda_R^*\cos{\frac{\theta}{2}}&-\lambda_R^*\cos{\frac{\theta}{2}}\\
\lambda_R^*\sin{\frac{\theta}{2}}&-\lambda_R^*\sin{\frac{\theta}{2}}\\
\lambda_R\sin{\frac{\theta}{2}}&-\lambda_R\sin{\frac{\theta}{2}}\\
\lambda_R\cos{\frac{\theta}{2}}&-\lambda_R\cos{\frac{\theta}{2}}
\end{pmatrix}   \,.
\end{gather}
\end{subequations}
The other two off-diagonal sub-matrices
are given by
$\mathcal{H}_{Ld}=\mathcal{H}_{dL}^\dag$
and $\mathcal{H}_{Rd}=\mathcal{H}_{dR}^\dag$.

\subsection{Josephson current }

From the description of the above low-energy effective Hamiltonian,
it seems that this system, which accommodates discrete energy levels,
can support only unitary evolution (quantum oscillations).
However, this is not true.
Note that the MFs at the ends of the nanowire are induced
via contact with superconductor
which has the Cooper pair reservoir.
Even under zero bias voltage, the both superconductors
(see Fig.~\ref{system}) can support stationary dc Josephson current,
if a phase difference between the superconductors is maintained.
This understanding allows us to apply the quantum transport
formalism of nonequilibrium Green's function (nGF)
to the present set-up with only discrete energy levels.

Following the nGF technique outlined in
Ref.\ [\onlinecite{Sun01,Sun02}],
the Josephson current reads
\begin{equation}
I=\frac{e}{h}\int \mathrm{d}\epsilon \, \mathrm{Re} \,
\mathrm{Tr}\{\tilde{\sigma}_z
[\tilde{\Sigma}(\epsilon) G_d(\epsilon)]^<\}  \,.
\end{equation}
This result is expressed in the Nambu representation.
Accordingly, $\tilde{\sigma}_z=diag\{1,1,-1,-1\}$,
which makes the Keldysh equation
$[\tilde{\Sigma}G_d]^<=\tilde{\Sigma}^< G_d^a + \tilde{\Sigma}^r G_d^<$
applicable in the compact form.
$G_d$ is the `reduced' effective Green's function
of the quantum dot, by accounting for the effect
of the `leads' (the Majoranas connected at both sides)
as self-energies (as usual in the nGF formalism for transport).  
Here we use $\tilde{\Sigma}$ to denote the difference of
the self-energies from the left and right wires (leads),
$\tilde{\Sigma}=\Sigma_L-\Sigma_R$.
The associated retarded and advanced self-energies
are given, respectively, by
\begin{equation}
\Sigma_{L/R}^{r(a)}(\epsilon)
=\mathcal{H}_{d\,L/R}\, g^{r(a)}_{L/R}(\epsilon)
\, \mathcal{H}_{L/R\,d}
\end{equation}
where $g^{r(a)}_{L/R}(\epsilon)
=[\epsilon-\mathcal{H}_{LL/RR}\pm i0^+]^{-1}$
are the retarded (advanced) Green's functions
of the isolated left/right wires.
For the lesser self-energies, one can apply:
$\Sigma^<_{L/R}(\epsilon)
=f(\epsilon)[\Sigma^a_{L/R}(\epsilon)
-\Sigma^r_{L/R}(\epsilon)]$,
where $f(\epsilon)$ is the Fermi-Dirac distribution function.
The retarded and advanced Green's functions read
$G_d^{r(a)}(\epsilon)=[\epsilon-{\cal H}_{dd}
-(\Sigma^{r(a)}_{L} + \Sigma^{r(a)}_{R})]^{-1}$,
and the lesser Green's function can be similarly obtained
by using
$G_d^<(\epsilon)=f(\epsilon)[G_d^a(\epsilon)-G_d^r(\epsilon)]$.
Using these relations together with some algebras, we obtain
\begin{eqnarray}
I&=&\frac{e}{h}\int \mathrm{d}\epsilon \mathrm{Re} \mathrm{Tr}[\tilde{\sigma}_z(\tilde{\Sigma}^a G_d^a
-\tilde{\Sigma}^r G_d^r)]f(\epsilon) \,.
\end{eqnarray}
This compact formula will be applied in this work to
calculate the Josephson current.

\section{Results}

In the numerical investigations, results will be calculated
under the influence of a couple of set-up parameters:
the gate voltage, which would affect the QD level $\varepsilon_d$;
the bent angle $\theta$ of the junction; the magnetic field in the QD area;
and the overlap strength of the Majorana wavefunctions ($\varepsilon_{L/R}$).
We would like to set $|\lambda_L|=|\lambda_R|=\lambda=1$
and scale all energies by $\lambda$.   
Also, we denote the Fermi level of the entire set-up
as reference (zero) energy,
the phase difference between the superconductors
as $\Delta\phi=\phi_R-\phi_L$,
and assume zero temperature and
identical nanowires ($\varepsilon_L=\varepsilon_R$).

\begin{figure}[htbp]
  \centering
  \includegraphics[scale=0.55]{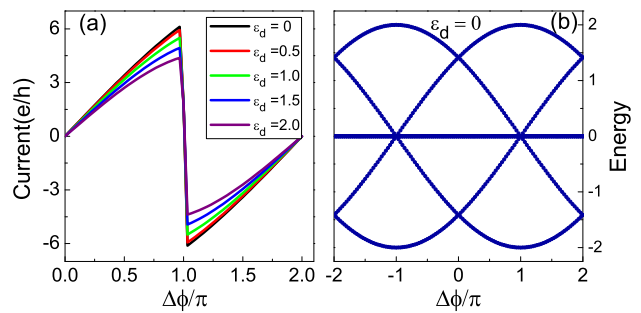}
  \caption{(color online)
(a) Josephson current {\it versus} the phase difference
between the two superconductors
for the straight Majorana--QD--Majorana junction ($\theta=0$),
under modulation of the QD energy level.
The parameters $\varepsilon_L=\varepsilon_R=0$
and the magnetic field ${\bf B}=0$.
(b) Example of the associated
in-gap energy diagram of the Josephson junction
in the topological phase, which can help us
understand the current behavior shown in (a). }\label{Fig2}
\end{figure}

\subsection{Effect of Dot Level Modulation}

We first display the result for the straight junction with $\theta=0$.
In this case, the unique Majorana feature is revealed,
as shown in Fig.~\ref{Fig2}(a),
by the robust ``jump" of the Josephson current at $\Delta\phi=\pi$
and the non-vanishing current
when modulating the dot level $\varepsilon_d$
(via modulation of the gate voltage applied).   
We see that, with the increase of $\varepsilon_d$
(even far away from the Fermi energy, i.e., $\varepsilon_d=0$),
the Josephson current only reduces by a small amount,
and the jump at $\Delta\phi=\pi$ always survives there,
being robust against the variation of $\varepsilon_d$.
For instance, for $\varepsilon_d=2.0$,
the amplitude of the Josephson current is only reduced
to about $70\%$ of the value at $\varepsilon_d=0$.

The both features revealed here are entirely different from
the normal superconductor--QD--superconductor junction.
In the normal (trivial) case,
the jump only appears at $\varepsilon_d=0$
and the current is to be strongly suppressed
when $\varepsilon_d$ deviates far away from the Fermi energy \cite{Sun02}.
We thus conclude that the features revealed in Fig.~\ref{Fig2}(a)
are closely associated with the superconductor-proximity-induced
nontrivial (topological) phase of the nanowire where the MF emerges.
In this case, the structure of the energy diagram
is as shown in Fig.~\ref{Fig2}(b), where we find the remarkable
zero-energy crossing points at $\Delta\phi=\pm\pi$ which are
responsible for the current ``jumps" as observed in Fig.~\ref{Fig2}(a).


Qualitatively speaking, the current is the sum of
all contributions from the occupied energy levels,
with each individual proportional to
the derivative of the associated energy curve
(with respect to $\Delta\phi$) \cite{Fu09}.
In this work, we calculate the current using Eq.\ (8).
We integrate the energy from $-\infty$ to the Fermi level,
implying that the system is always in thermal equilibrium
especially when $\Delta\phi$ passes through $\pi$.
This treatment corresponds to certain relaxation mechanism
involved, as a consequence of {\it particle addition/loss}
from/to the surrounding environment.
As a consequence, we obtain the $2\pi$ periodic current
(as a function of $\Delta\phi$), Fig.~\ref{Fig2}(a).
Only under the fermion-number-parity conservation,
the so-called $4\pi$ periodic current can be expected.
We will address this issue later in more detail. \\
\\


\subsection{Effect of Magnetic Fields}

We now consider the possible effects of magnetic field in the dot area.
For the Zeeman splitting of the dot level
caused by the magnetic field along the $z$ axis,
we can understand that the effect is the same as
the electric gate-modulation of the dot level,
as shown in Fig.~\ref{Fig2}(a).

\begin{figure}[H]
  \centering
  \includegraphics[scale=0.75]{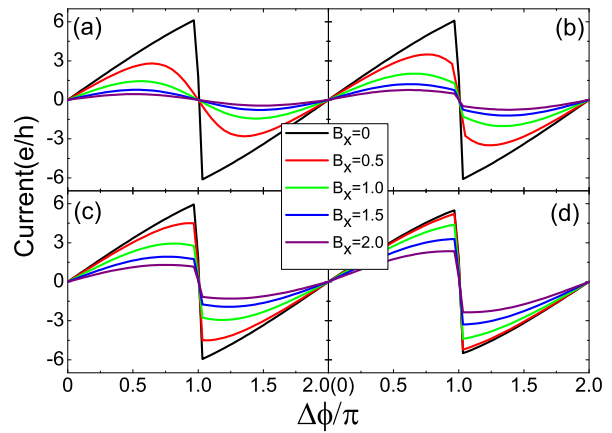}
  \caption{(color online)
Current-phase behavior under the influence of
the $x$-component of the magnetic field in the dot area.
Results for several locations of the dot level are displayed:
(a) $\varepsilon_d=0$; (b)$\varepsilon_d=0.2$;
(c) $\varepsilon_d=0.5$; and (d) $\varepsilon_d=1.0$. }\label{Fig3}
\end{figure}

The effect of $B_x$ (magnetic field along the $x$ axis)
is shown in Fig.~\ref{Fig3}.
For the ideal configuration with $\varepsilon_d=0$,
the jump at $\Delta\phi=\pi$ disappears and evolves
to a rounded transition behavior.
And, with the increase of $B_x$, the Josephson current
quickly decreases and vanishes at last.

For $\varepsilon_d\neq 0$, similar modulation effect of $B_x$
on the amplitude of the Josephson current
is caused by the spin-selective coupling between the Majorana and QD states,
by noting that the magnetic field $B_x$ has a role of rotating the electron spin,
leading thus to a change of the $z$-component of the spin.
However, in this more ``relaxed" case ($\varepsilon_d\neq 0$),
the jump behavior at $\Delta\phi=\pi$ survives.
This indicates that the Majorana characteristic {\it jump}
behavior is robust against the
deviation of the dot level from the Fermi energy,
even under the influence of magnetic field in the QD area.

Moreover, out of simple expectation,
if the dot level $\varepsilon_d$ is far away
from the Fermi energy (e.g. $\varepsilon_d=1$),
the Josephson current is reduced by $B_x$
much more modestly,
than in the case of $\varepsilon_d=0$,
see, e.g., the results in Fig.~\ref{Fig3}(a) and (d).
Indeed, this behavior is very counterintuitive,
since in usual case the largest current
always appears at resonant tunneling \cite{add7,add8},
i.e., the dot level $\varepsilon_d$ at the Fermi energy. \\
\\


\subsection{Effect of Junction Bending}

\begin{figure}[htbp]
  \centering
  \includegraphics[scale=0.75]{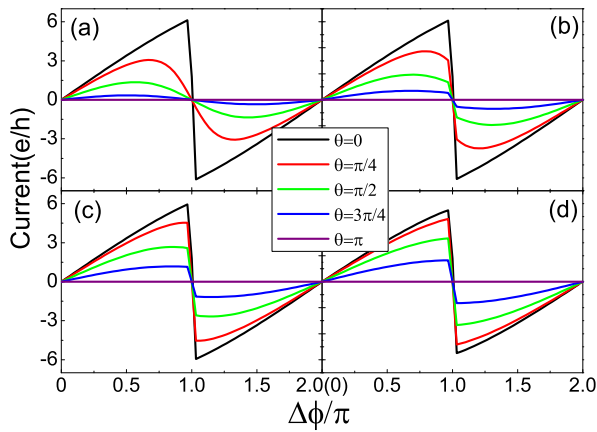}\\
  \caption{(color online)
Effect of junction bending on the current-phase behavior.
Displayed are results for several locations of the dot level:
(a) $\varepsilon_d=0$; (b)$\varepsilon_d=0.2$;
(c) $\varepsilon_d=0.5$; and (d) $\varepsilon_d=1.0$. }\label{Fig4}
\end{figure}

As remarked in the Introduction,
the spin-selective coupling can be manifested also
by bending the junction, i.e.,
by altering the mutual angle ($\theta$) between the nanowires.
The results are displayed in Fig.~\ref{Fig4},
which actually resemble what we observed in Fig.~\ref{Fig3}.
That is, the ``jump" at $\Delta\phi=\pi$ is rounded
in Fig.~\ref{Fig4}(a) for $\varepsilon_d=0$,
while in Fig.~\ref{Fig4}(b), (c), and (d) for $\varepsilon_d\neq 0$,
the ``jump" survives and the maximum (amplitude) of the current
decreases with the increase of the bent angle $\theta$.
Again, out of expectation, for finite $\theta$
(e.g. $\theta=\frac{\pi}{4}$, $\frac{\pi}{2}$ and $\frac{3\pi}{4}$),
the amplitude of the Josephson current
would increase with the deviation of the dot level
from the Fermi energy.
This means that, for a bent Majorana-Josephson junction,
the stronger the resonance condition is violated,
a larger supercurrent will flow through the junction.
We finally mention that, for all cases,
the parameter $\theta=\pi$ simply means
rotating the right wire to the same side of the quantum dot
and in parallel to the left wire,
which would result in a completely vanished current.

\subsection{Energy Diagram Based Interpretation}

\begin{figure}[!b]
\centering
\includegraphics[scale=0.75]{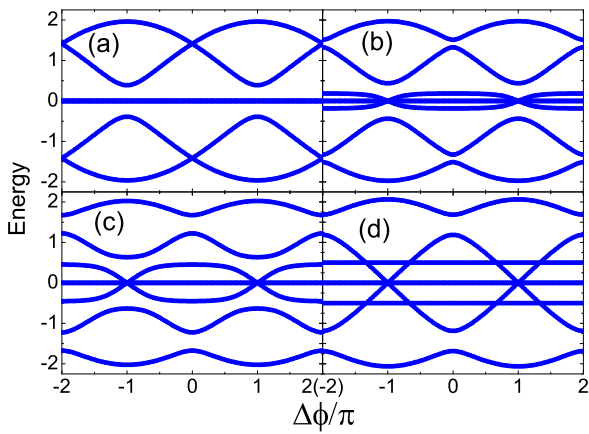}
\caption{(color online)
Energy diagram of the junction for a couple of set-up parameters:
(a) $\varepsilon_d=0, ~\theta=0.25\pi$;
(b) $\varepsilon_d=0.2, ~\theta=0.25\pi$;
(c) $\varepsilon_d=0.5, ~\theta=0.25\pi$;
and (d) $\varepsilon_d=0.5, ~\theta=0$.  }\label{Fig5}
\end{figure}

We found through Figs.~\ref{Fig2},~\ref{Fig3} and~\ref{Fig4} that,
regardless of the magnetic field ${\bf B}$
and the mutual angle $\theta$,
the current jump can safely survive
and the amplitude of the current
maintains a large value,
if the dot level $\varepsilon_d$ violates
the resonance condition.
We may further understand the behaviors as follows,
with the help of the energy diagrams
of Fig.~\ref{Fig2}(b) and Fig.~\ref{Fig5}.

First, in Fig.~\ref{Fig2}(b) for $\varepsilon_d=0$ and $\theta=0$,
the flat zero-energy level are of four-fold degeneracy,
i.e., for the states $\gamma_{L2}$,
$\gamma_{R1}$, $d_{\downarrow}$ and $d^{\dagger}_{\downarrow}$.
The other four phase-difference ($\Delta\phi$) dependent
eigen-energies are from the coupling of the states $\gamma_{L1}$,
$\gamma_{R2}$, $d_{\uparrow}$ and $d^{\dagger}_{\uparrow}$.
To be more specific,
the two Majoranas $\gamma_{L1}$ and $\gamma_{R2}$
couple commonly to the spin-up dot state,
resulting in the `crossing' structure
at zero energy at $\Delta\phi=\pm\pi$,
which is similar to the result of
direct coupling of two Majoranas.\cite{pei1}
It is just owing to this zero-energy crossing
(at the Fermi energy)
that the large Josephson current with abrupt jump
is resulted in,
in the presence of `relaxation' or thermal equilibrium.

Second, for the dot level $\varepsilon_d\neq 0$
but $\theta=0$ [see Fig.~\ref{Fig5}(d)],
the Majorana states $\gamma_{L2}$ and $\gamma_{R1}$
have also the flat zero-energy (independent of $\Delta\phi$);
however, the energies of
$d_{\downarrow}$ and $d^{\dagger}_{\downarrow}$
move to $\varepsilon_d$ and $-\varepsilon_d$, respectively.
These would lead to an opening of the high-energy crossing
at $\Delta\phi=0$ (or $\pm 2\pi$),
but still keeping the zero-energy crossings at $\Delta\phi=\pm\pi$
caused by $\gamma_{L1}$ and $\gamma_{R2}$.
As a result, the Josephson current keeps a large value
and the jump survives, because the current are dominantly
contributed by the zero-energy crossing states.

Third, if $\theta\not=0$
and $\varepsilon_d=0$ [see Fig.~\ref{Fig5}(a)],
we see that the zero-energy degeneracy of $\gamma_{L1}$
and $\gamma_{R2}$ at $\Delta\phi=\pm\pi$ is removed.
The disappearance of the zero-energy crossings
leads to a strong reduction of the Josephson current
and rounding the jump to smooth transition,
as shown in Fig.~\ref{Fig4}(a).

Finally, for the case of both $\theta\not=0$
and $\varepsilon_d\not=0$,
as shown in Fig.~\ref{Fig5}(b) and (c),
the absence of efficient energy level interaction
between the Majorana and QD states
(owing to $\varepsilon_d\neq 0$)
does not remove the zero-energy degeneracy of
$\gamma_{L1}$ and $\gamma_{R2}$ at $\Delta\phi=\pm\pi$.
Then, the zero-energy crossing structure of the energy spectrum
at $\Delta\phi=\pm\pi$
results in a large Josephson current with abrupt jumps,
despite that the dot level $\varepsilon_d$ deviates far
from the Fermi energy.


\subsection{Effect of Majorana Interaction}

\begin{figure}[H]
  \centering
  \includegraphics[scale=0.75]{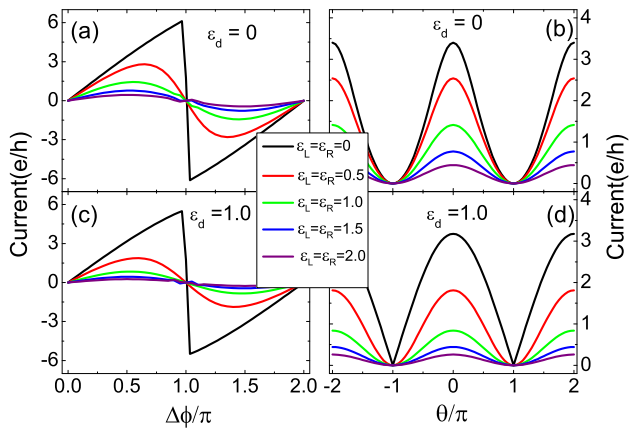}\\
  \caption{(color online)
(a) and (c):
Effect of Majorana interaction
(in the same nanowire manifested by nonzero $\varepsilon_{L,R})$,
which would reduce the Josephson current and destroy
the jump behavior at $\Delta\phi=\pm\pi$.
(b) and (d): Nanowire-orientation-angle dependence of the current at $\Delta\phi=0.5\pi$.
Results for dot level $\varepsilon_d=0$ and $\varepsilon_d=1$
are presented, respectively.
}\label{Fig6}
\end{figure}

Below we show that, in order to observe
the featured behaviors discussed above,
the overlap of the Majorana wavefunctions
at the ends of the same nanowire
should be negligibly small.
Or, equivalently, the Majorana interaction
should be negligibly small.
In Fig.~\ref{Fig6} we display the result for
$\varepsilon_L=\varepsilon_R\neq 0$,
which corresponds to nonzero coupling
between the two Majoranas in the same nanowire.
Indeed, we find that the amplitude of
the Josephson current is strongly reduced
and the jump disappears,
with the increase of $\varepsilon_{L,R}$
[see Fig.~\ref{Fig6}(a) and (c)].
The basic reason is that the Majorana interaction
in the same nanowire destroys the zero-energy Majorana state.

We have also checked that the large Josephson current
with jump behavior cannot be restored
by altering the dot level,
even with $\varepsilon_d$ in resonance
with $\varepsilon_L$ and $\varepsilon_R$ [see Fig.~\ref{Fig6}(c) and (d)].
The reason is that, if $\varepsilon_L=\varepsilon_R\neq0$,
the zero-energy Majorana states $\gamma_{L1}$ and $\gamma_{R2}$
are destroyed and the zero-energy crossings
at $\Delta\phi=\pm\pi$ disappear.
As a consequence, the jumps at $\Delta\phi=\pm\pi$
are replaced by rounded transitions,
and the current is strongly reduced.

In Fig.~\ref{Fig6}(b) and (d), taking the current at $\Delta\phi=\pi/2$,
we show again the $\theta$ dependence of current,
in the presence of Majorana coupling ($\varepsilon_{L,R}\neq 0$).
From this result we see clearly that,
in order to obtain a larger super-current,
we should make the nanowire longer
than the superconductor coherence length
to ensure the emergence of Majorana zero modes
at the ends of the nanowire.

\subsection{$4\pi$ Periodic Current}

So far we have assumed that the system
always relaxes to a thermal equilibrium
when we vary the phase difference $\Delta\phi$.
In particular, at the zero-energy crossings at $\Delta\phi=\pm\pi$,
this relaxation is accompanied by either an addition
or a loss of a single particle,
which therefore changes the parity of the particle numbers
(i.e., the fermion parity).
If such relaxation channel is blocked
or the fermion parity is conserved,
rather than the $2\pi$ periodic current we obtained above,
a remarkable $4\pi$ periodic Josephson current can be expected,
which is usually regarded as
one of the most prominent Majorana signatures.

The $4\pi$ periodic current can be calculated
as well by using Eq.\ (8),
based on the following technique:
When we increase $\Delta\phi$ after passing through $\pm\pi$,
for the state occupation of the levels
crossing at zero energy (the Fermi level) at $\Delta\phi=\pm\pi$,
we replace the occupation of the lower level
under the Fermi energy
by its counterpart above the Fermi level
(owing to the fermion parity conservation),
while satisfying the condition of thermal equilibrium
after this replacement.

\begin{figure}[h]
  \centering
  \includegraphics[scale=0.75]{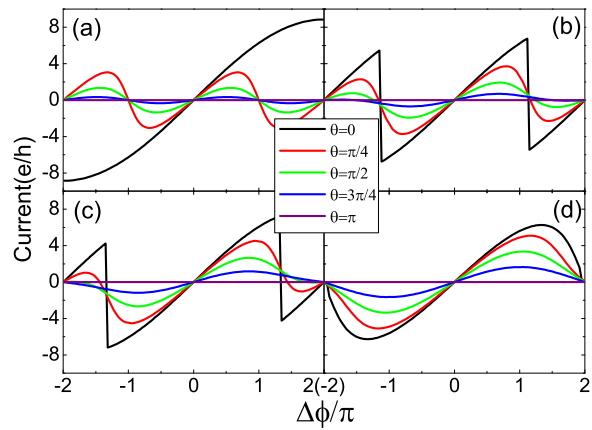}\\
  \caption{(color online)
  The so-called
  $4\pi$ periodic current for a couple of set-up parameters, i.e., the
  nanowire-orientation-angles (as shown by the inset) and the dot levels
  (a) $\varepsilon_d=0$, (b) $\varepsilon_d=0.2$,
  (c) $\varepsilon_d=0.5$, and (d) $\varepsilon_d=1.0$.   }\label{Fig7}
\end{figure}

The results of the $4\pi$ periodic current are shown
in Fig.~\ref{Fig7}, for one period.
Compared with the $2\pi$ periodic current,
we find that the jumps at $\Delta\phi=\pm\pi$
disappear for all the $4\pi$ periodic currents.
However, for the case of $\theta=0$ and $\varepsilon_d\neq 0$
similar jumps may appear {\it near}
(but not {\it at}) $\Delta\phi=\pm\pi$,
as observed in Fig.~\ref{Fig7}(b) and (c),
owing to the accidental energy crossings at the specific phases.
Again, along the increase of the angle $\theta$,
the amplitude of the current decreases.
However, the current is more strongly suppressed
in the case of $\varepsilon_d=0$,
while for larger deviation of $\varepsilon_d$
(from the Fermi level) the current is less reduced.

\section{Summary}\label{Conclusions}

To summarize, we have investigated the dc Josephson supercurrent
through the Majorana--quantum dot--Majorana junction.
Our particular interest is the consequence of the unique
spin-selective coupling between the Majorana and dot states,
which emerges only in the topological phase
and will drastically influence the current
through bent junctions and/or
in the presence of magnetic fields in the dot area.
Differing from the typical resonant tunneling
behavior of the supercurrent
through similar system in normal phase such as
the superconductor--quantum dot--superconductor junction,
we uncovered some counterintuitive results
associated with the exotic nature of the Majorana fermion.

For instance, even for a straight junction
and without magnetic field in the dot area,
when the dot level deviates considerably
from the Fermi energy,
the Josephson supercurrent keeps a large
amplitude of oscillation with
the superconductor phase difference $\Delta\phi$
and reveals abrupt jumps of current at $\Delta\phi=\pm \pi$.
Drastically, this result differs from the usual
resonant tunneling behavior through similar system in normal phase.
For a bent junction and/or in the presence of magnetic field in the dot,
richer unexpected behaviors are found.
In resonance (the dot level aligned with the Fermi energy),
we find that the supercurrent is to be strongly reduced by
either the junction bending or the magnetic fields in the dot.
At the same time, the current jumps at $\Delta\phi=\pm \pi$ are rounded.
However, if the dot level deviates from the Fermi energy
(i.e., violates the resonant tunneling condition),
the supercurrent can, on the contrary,
maintain a large amplitude of current and
the current jumps robustly survive at $\Delta\phi=\pm \pi$,
even under
the influence of junction bending and magnetic fields in the dot.
We expect these findings to be useful
in future design of novel circuit devices
based on quantum dots and Majorana nanowires.

\section*{Acknowledgements}

This work was supported by NBRP of China (Grant No. 2015CB921102),
NSF-China under Grants No. 11675016, No. 11274364 and No. 11574007,
the Beijing Natural Science Foundation under No. 1164014, the China Postdoctoral Science Foundation funded project (Grant No. 2016M591103)
and the Fundamental Research Funds for the Central Universities.

\end{document}